\documentclass[apj,iop]{emulateapj}

\usepackage[colorlinks=true,citecolor=blue,linkcolor=cyan,breaklinks]{hyperref}
\usepackage{ amssymb }
\usepackage{subfigure}
\usepackage{longtable}

\shorttitle{Mass and Light of Abell 370}
\shortauthors{Strait et al.}

\begin{document}

\title{Mass and Light of Abell 370: A Strong and Weak Lensing Analysis}

\author{V. Strait\altaffilmark{1}}
\author{M. Brada{\v c}\altaffilmark{1}}
\author{A. Hoag\altaffilmark{1}}
\author{K.-H. Huang\altaffilmark{1}}
\author{T. Treu\altaffilmark{2}}
\author{X. Wang\altaffilmark{2,4}}
\author{R. Amorin\altaffilmark{6,7}}
\author{M. Castellano\altaffilmark{5}}
\author{A. Fontana\altaffilmark{5}}
\author{B.-C. Lemaux\altaffilmark{1}}
\author{E. Merlin\altaffilmark{5}}
\author{K.B. Schmidt\altaffilmark{3}}
\author{T. Schrabback\altaffilmark{8}}
\author{A. Tomczack\altaffilmark{1}}
\author{M. Trenti\altaffilmark{9,10}}
\author{B. Vulcani\altaffilmark{9,11}}
\affil{\altaffilmark{1}Physics Department, University of California, Davis, CA 95616, USA}
\affil{\altaffilmark{2}Department of Physics and Astronomy, UCLA, Los Angeles, CA, 90095-1547, USA}
\affil{\altaffilmark{3}Leibniz-Institut f{\"u}r Astrophysik Postdam (AIP), An der Sternwarte 16, 14482 Potsdam, Germany}
\affil{\altaffilmark{4}Department of Physics, University of California, Santa Barbara, CA, 93106-9530, USA}
\affil{\altaffilmark{5}INAF - Osservatorio Astronomico di Roma Via Frascati 33 - 00040 Monte Porzio Catone, 00040 Rome, Italy}
\affil{\altaffilmark{6}Cavendish Laboratory, University of Cambridge, 19 JJ Thomson Avenue, CB3 0HE, Cambridge, UK}
\affil{\altaffilmark{7}Kavli Institute for Cosmology, University of Cambridge, Madingley Rd., CB3 0HA, Cambridge, UK}
\affil{\altaffilmark{8}Argelander-Institut f{\"u}r Astronomie, Auf dem H{\"u}gel 71, D-53121 Bonn, Germany}
\affil{\altaffilmark{9}School of Physics, University of Melbourne, Parkville, Victoria, Australia}
\affil{\altaffilmark{10}ARC Centre of Excellence fot All Sky Astrophysics in 3 Dimensions (ASTRO 3D)}
\affil{\altaffilmark{11}INAF - Astronomical Observatory of Padora, 35122 Padova, Italy}

\begin{abstract}
We present a new gravitational lens model of the Hubble Frontier Fields cluster Abell 370 (\mbox{$z = 0.375$}) using imaging and spectroscopy from \emph{Hubble Space Telescope} and ground-based spectroscopy. We combine constraints from a catalog of 909 weakly lensed galaxies and 39 multiply-imaged sources comprised of 114 multiple images, including a system of multiply-imaged candidates at \mbox{$z=7.84\pm0.02$}, to obtain a best-fit mass distribution using the cluster lens modeling code Strong and Weak Lensing United. As the only analysis of A370 using strong and weak lensing constraints from Hubble Frontier Fields data, our method provides an independent check of assumptions on the mass distribution used in other methods. Convergence, shear, and magnification maps are made publicly available through the HFF website\footnote{http://www.stsci.edu/hst/campaigns/frontier-fields}. We find that the model we produce is similar to models produced by other groups, with some exceptions due to the differences in lensing code methodology. In an effort to study how our total projected mass distribution traces light, we measure the stellar mass density distribution using Spitzer/Infrared Array Camera imaging. Comparing our total mass density to our stellar mass density in a radius of 0.3 Mpc, we find a mean projected stellar to total mass ratio of \mbox{$\langle f* \rangle = 0.011 \pm 0.003$} (stat.) using the diet Salpeter initial mass function. This value is in general agreement with independent measurements of \mbox{$\langle f* \rangle$} in clusters of similar total mass and redshift.
\end{abstract}


\section{Introduction}

Cluster lens modeling has been used for decades as a tool to retrieve intrinsic properties of lensed sources for various types of scientific study. For example, investigation into properties of high redshift (\mbox{$z > 6$}) galaxies plays an essential role in understanding early galaxy evolution and the reionization of the universe. By measuring the number counts of high-redshift sources as a function of magnitude, we can obtain the ultraviolet luminosity function (UV LF), which allows us to infer properties such as star formation rate density, an essential piece to understanding the role that galaxies played in the reionization of the universe (e.g., \citealp{sch14b,fink15,mas15,rob15,cast16b,liv17,bou17,ish18}). At lower redshifts (\mbox{$z = 0.7-2.3$}), it is possible to measure spatially resolved kinematics and chemical abundances for the brightest sources (e.g., \citealp{chri12,jon15,wan15,vul16,lee16,mas17,gir18,pat18}). Probing the faint end of both of these samples is challenging with the detection limits associated with blank fields. Using the gravitational lensing power of massive galaxy clusters, fainter sources can be studied in greater detail. To infer many of these sources' intrinsic properties (e.g., star formation rate and stellar mass), magnification maps are needed. Using strongly lensed sources and weakly lensed galaxies as constraints, lens models produce magnification and mass density maps. 

Abell 370 (\mbox{$z = 0.375$}, A370 hereafter) was the first massive galaxy cluster observed for the purposes of gravitational lensing, initially alluded to by \cite{lyn86} with follow-up by \cite{lyn89}. The cluster was also studied in depth by \cite{sou87,ham89} because of the giant luminous arc in the south, which led to the first lens model of A370 by \cite{ham87}. \cite{rich10} provided one of the first strong lensing models of A370 which used data from HST imaging campaigns of the cluster, and weak lensing analyses followed soon after \citep{ume11,med11}. Since then, deeper imaging data have been taken of A370 by the Hubble Frontier Fields program (HFF: PI Lotz \#13495, \citealp{lotz17}), an exploration of six massive galaxy clusters selected to be among the strongest lenses observed to date. Spectroscopic campaigns such as the Grism Lens-Amplified Survey from Space (GLASS) (\citealp{sch14b,tre15}) and the Multi-Unit Spectroscopic Explorer Guaranteed Time Observations (MUSE GTO, \citealp{lag17}) have obtained spectroscopic redshifts for nearly all of the strongly lensed systems discovered by HFF data. In this work we use 37 spectroscopically confirmed strongly lensed background galaxies and 2 with robust photometric redshifts, totalling 39 systems, as constraints for our lens model (see Section \ref{mi_section} for details). It has been shown by e.g., \cite{john16} that the most important parameter in constraining cluster lens models is the fraction of high quality (i.e. spectroscopically confirmed) multiply imaged systems to total number of systems. The 37 spectroscopically confirmed strongly lensed systems out of a total 39 combined with high quality weak lensing data in A370 has led to some of the most robust lens models of any cluster to date. 


Several modeling techniques have been used to make magnification and total mass density maps of A370 (\citealp{rich14,john14,kaw17,lag17,die18}), each making various assumptions about the mass distribution. For example, \cite{rich14,john14,kaw17} produce high resolution maps using parametric codes to constrain the mass distribution using a simple Bayesian parameter minimization and an assumption that mass traces light. Other techniques use adaptive grid models, such as \cite{die18} and the model presented here (although \cite{die18} assumes that mass traces light and our method does not do so beyond the choice of initial model). These have the potential to test for systematic errors that arise from assumptions about the mass distribution. A robust measurement of error using a range of magnification maps becomes essential for any measurement made at high magnification (\mbox{$\mu > 20$}). For example, \cite{bou17} showed that at the faint end of the UV LF (which cannot yet be probed without lensing and where sources are more likely highly magnified), magnification errors become large. In addition, \cite{men17} has shown that the error in magnification is proportional to magnification. In response to the need for a wide range of lens models for each cluster being studied, magnification maps from several teams, including ours, are publicly available on the Hubble Frontier Fields website.\footnote{http://www.stsci.edu/hst/campaigns/frontier-fields} 



While our method does not produce the highest resolution maps, we include both strong and weak lensing constraints and, apart from the initial model, make no assumptions about total mass distribution. Stellar mass can be independently measured using the observed stellar light, allowing total mass density maps of galaxy clusters to be compared to stellar mass density in order to obtain a stellar mass density to total mass density ratio (\mbox{$f\text{*}$}). This provides an independent way to see how light does or does not trace total mass in our model. In this paper, we present magnification, convergence, stellar mass density and \mbox{$f\text{*}$} maps of A370. 

The structure of the paper is as follows: we present a description of imaging and spectroscopic data in Section 2, a description of our gravitational lens modeling code, constraints we use in our model, and a description of our stellar mass measurement in Section 3. Following this, we present a stellar mass density to total mass density map in Section 4 and conclude in Section 5. Throughout the paper, we will give magnitudes in the AB system (Oke et al. 1974), and we assume a $\Lambda$CDM cosmology with \mbox{$h = 0.7$}, \mbox{$\Omega_m = 0.3$}, and \mbox{$\Omega_{\Lambda} = 0.7$}.
\begin{figure*}[ht!!!]
    \centering
    \includegraphics[width=14cm]{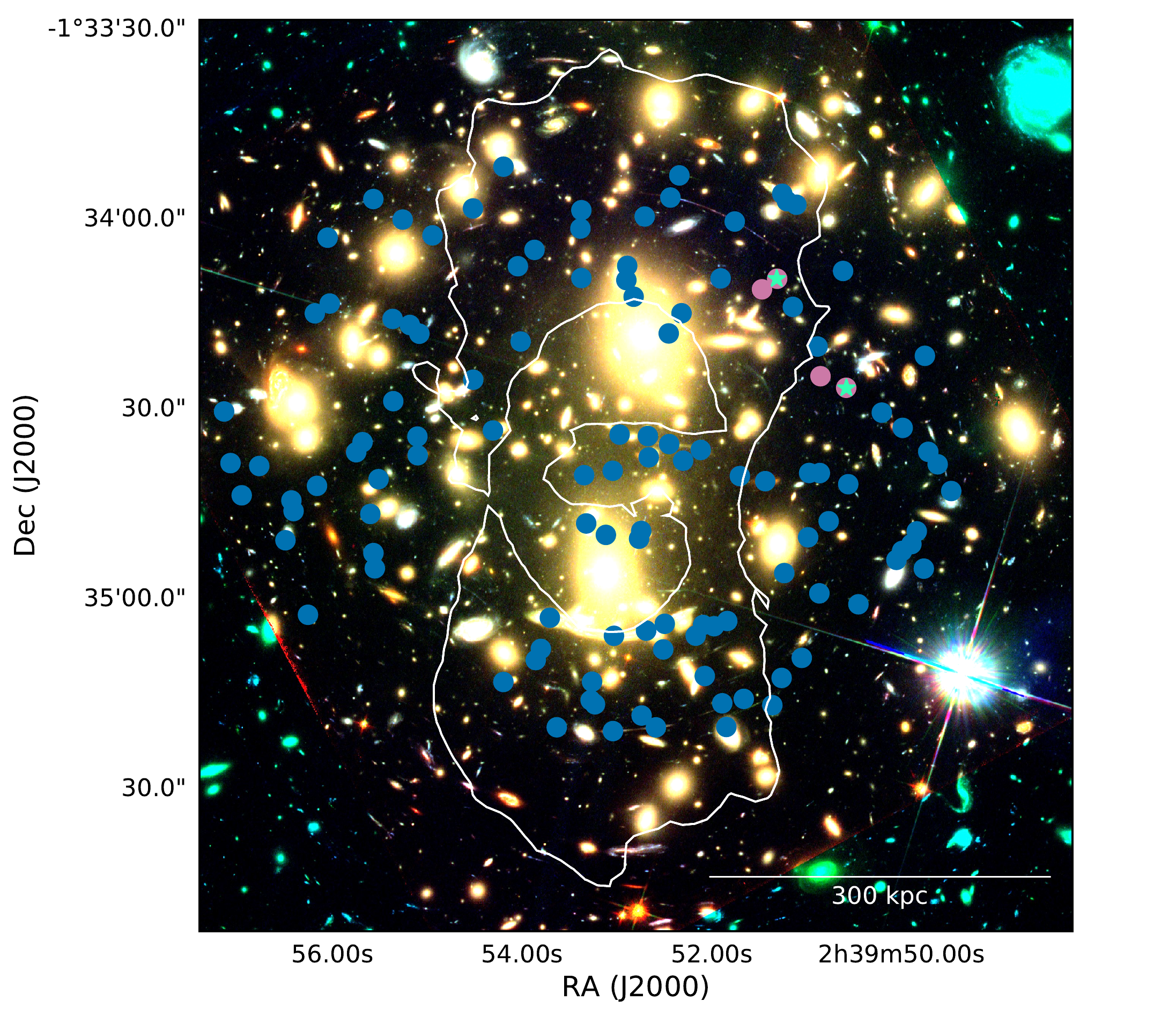}
    \caption{The critical curve at \mbox{$z=7.84$}, the redshift of multiply imaged system 11, for our model with multiple images marked as circles. Blue circles correspond to systems with spectroscopic redshifts (more secure) and magenta circles have photometric redshifts. We show the multiple images in system 11 as cyan stars. The color image is a combination of HST filters: F105W, F606W, F814W. The orientation is north up, east to the left.}
    \label{fig:cc_comp}
\end{figure*}
\begin{figure}[h!]
    \begin{subfigure}
        \centering
        \includegraphics[width=9cm]{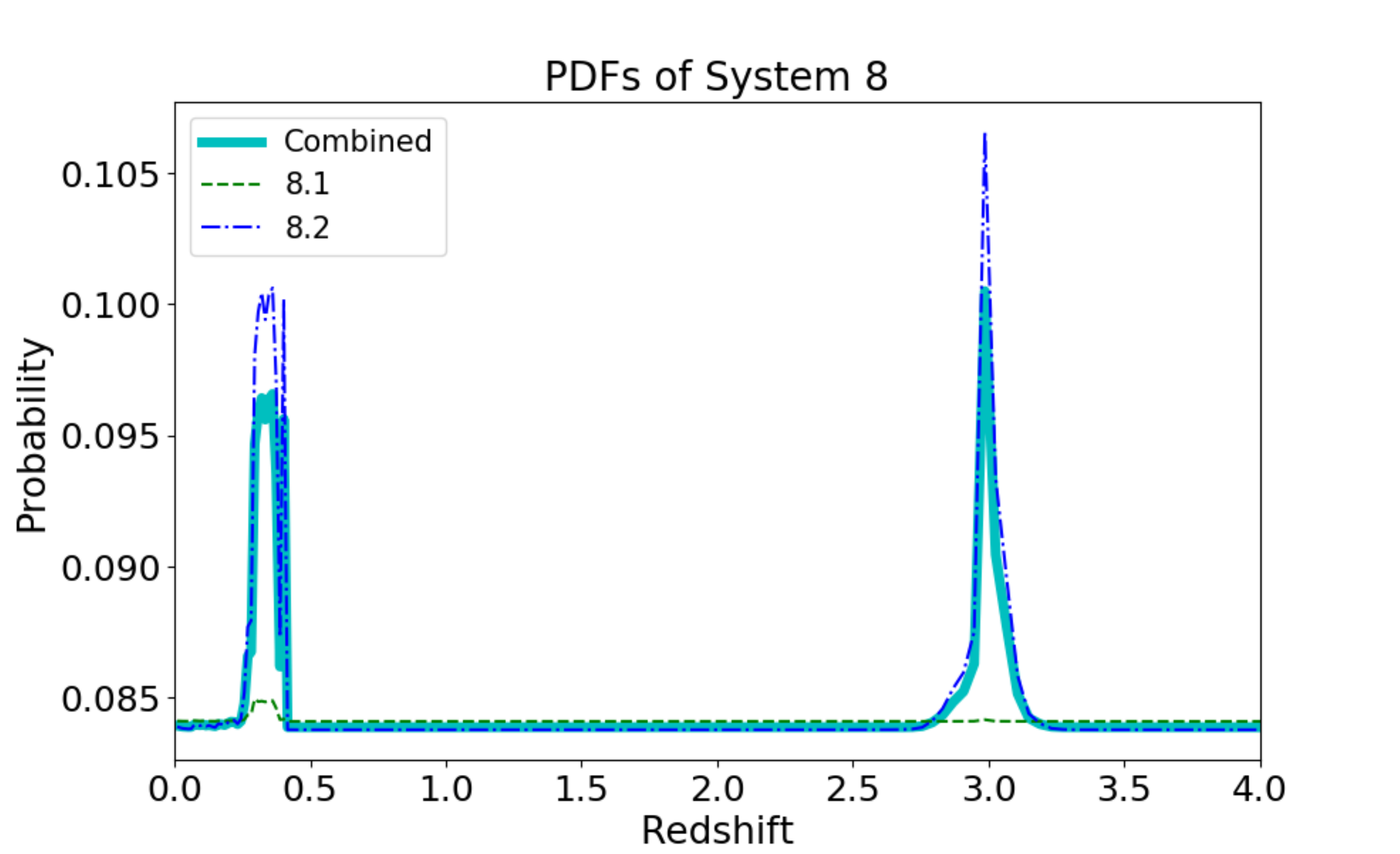}
    \end{subfigure}
    \begin{subfigure}
        \centering
        \includegraphics[width=9cm]{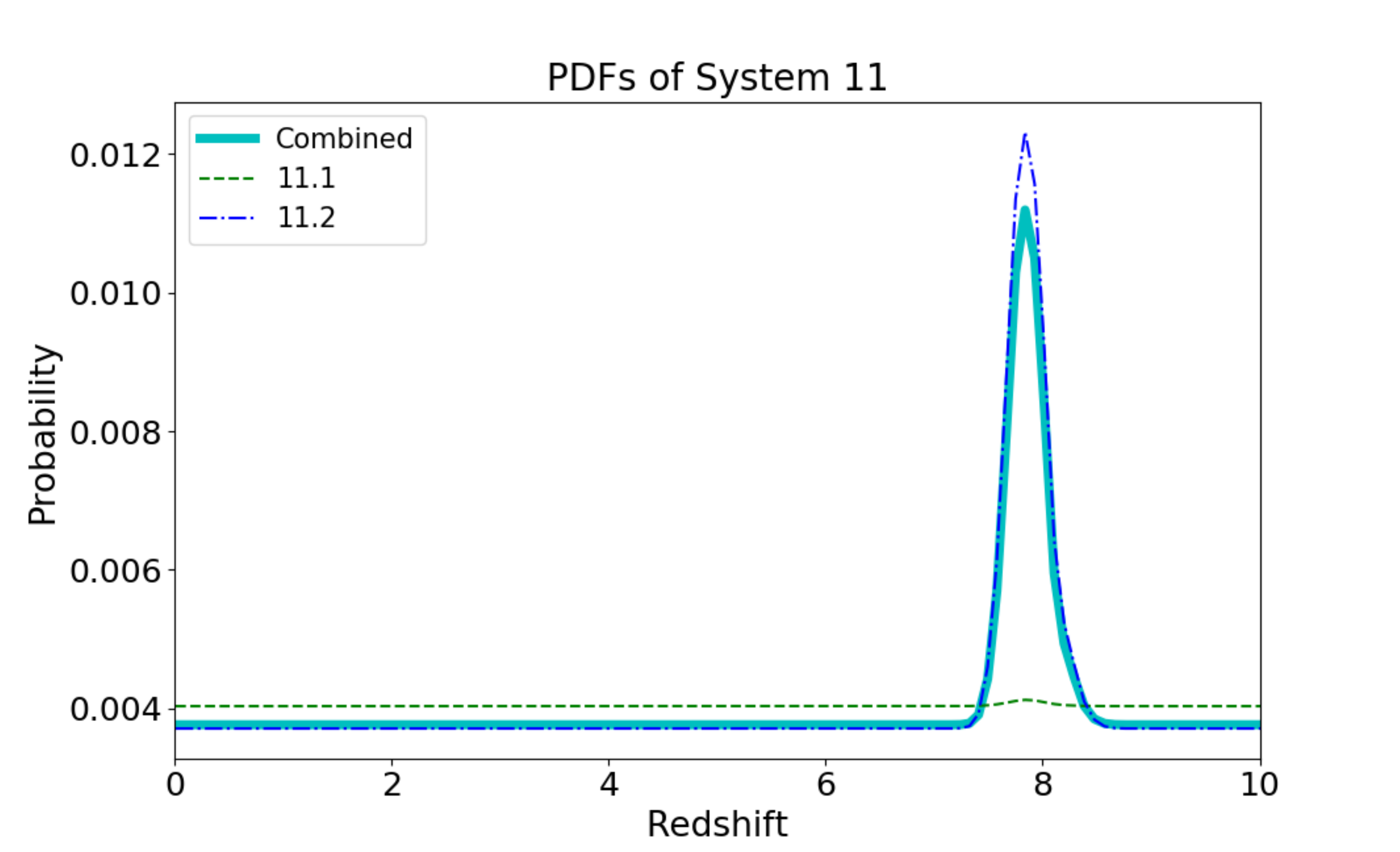}
    \end{subfigure}
\caption{Normalized probability distribution functions for multiple image systems 8 (top) and 11 (bottom), the two systems without a spectroscopic redshift. Image 8.1 is most likely contaminated by cluster members, and shows  a peak near the cluster redshift. However, there is a small peak at \mbox{$z\sim3$}. Because the multiple images in systems 8 and 11 have similar morphologies, surface brightnesses, and colors, we use all images as constraints in our model. Peak redshifts and 68\% confidence intervals are listed in Table \ref{tbl-2}.}
\label{811}
\end{figure}
\section{Observations and Data}
\subsection{Imaging and Photometry}
A combination of programs from \emph{Hubble Space Telescope} (HST), \emph{Very Large Telescope}/High Acuity Wide-field K-band Imager (VLT/HAWK-I), and \emph{Spitzer}/InfraRed Array Camera (IRAC) contribute to the broadband flux density measurements used in this paper. HST imaging is from HFF as well as a collection of other surveys (PI E. Hu \#11108, PI K. Noll \#11507, PI J.-P, Kneib \#11591, PI T. Treu \#13459, PI R. Kirshner \#14216),  and consists of deep imaging from the Advanced Camera for Surveys (ACS) in F435W (20 orbits), F606W (10 orbits), and F814W (52 orbits) and images from the Wide Field Camera 3 (WFC3) in F105W (25 orbits), F140W (12 orbits), and F160W (28 orbits). These images were taken from the Mikulski Archive for Space Telescope\footnote{https://archive.stsci.edu/} and were also used for visual inspection of multiply-imaged systems. 

Ultra-deep Spitzer/Infrared Array Camera (IRAC) images come from Spitzer Frontier Fields (PI T. Soifer, P. Capak, \citealp{lotz17}, Capak et al. in prep.)\footnote{http://irsa.ipac.caltech.edu/data/SPITZER/Frontier/} and are used for photometry and creation of a stellar mass map. These images reach 1000 hours of total exposure time of the 6 Frontier Fields clusters and parallel fields in each IRAC channel. All reduction of the Spitzer data follows the routines of \cite{hua16}. In addition to HST and Spitzer, we use data from K-band Imaging of the Frontier Fields (``KIFF", \citealp{bram16}), taken on VLT/HAWK-I, reaching a total of 28.3 hours exposure time for A370. 

\subsection{Spectroscopy}
Spectra are obtained from a combination of MUSE GTO observations and the Grism Lens Amplified Survey from Space (GLASS, PI Treu, HST-GO-13459, \citealp{sch14b,tre15}), and are used for obtaining secure redshifts for our strong lensing constraints. Spectroscopic redshifts for systems 1-4, 6, and 9 in Table \ref{tbl-2} are provided by \cite{die18}, which were originally obtained from GLASS\footnote{http://glass.astro.ucla.edu/} spectra. Data from MUSE (\citealp{lag17}; Lagattuta et al., in prep), provide confirmations of these as well as 31 other systems, totalling 39 spectroscopically confirmed systems, consisting of 114 multiple images. These are listed in Table \ref{tbl-2}, along with quality flags as defined in \cite{die18}. These range from 4 (best, determined by multiple high S/N emission lines) to 1 (worst, determined by one tentative, low S/N feature). The vast majority of the systems used in this work are quality flag (QF) 3, with only one image in one system with QF=1 (see Section \ref{mi_section} for details). 

\section{Analysis}
\subsection{Photometry}\label{phot_section}
For photometry of systems that do not have any spectroscopic constraints, we follow the procedure for the ASTRODEEP catalogs described by \cite{merl16,cast16a,dic17}. Using the seven HFF wideband filters (F435W, F606W, F814W, F105W, F125W, F140W, F160W), HAWK-I K-band imaging \citep{bram16}, and Spitzer/IRAC [3.6] and [4.5] channels (Capak et al., in prep.), the ASTRODEEP catalogs include subtraction of  intracluster light (ICL) and the brightest foreground galaxies from the images. ICL subtraction is done using T-PHOT \citep{merl15}, designed to perform PSF-matched, prior-based, multi-wavelength photometry as described in \cite{merl15,merl16}. This is done by convolving cutouts from a high resolution image (in this case, F160W) using a low resolution PSF transformation kernel that matches the F160W resolution to the IRAC (low-resolution) image. T-PHOT then fits a template to each source detected in F160W to best match the pixel values in the IRAC image. 

After all fluxes are extracted, colors in HST and IRAC images are calculated and used to estimate a probability density function (PDF) for each source using the redshift estimation code Easy and Accurate Redshifts from Yale (EAZY, \citealp{bram08}), which compares the observed SEDs to a set of stellar population templates. Using linear combinations of a base set of templates from \citeauthor{bruz03} (\citeyear{bruz03}, BC03), EAZY performs \mbox{$\chi^2$} minimization on a user-defined redshift grid, in our case ranging from \mbox{$z=0.1-12$} in linear steps of \mbox{$\delta z=0.1$}, and computes a PDF from the minimized \mbox{$\chi^2$} values.

To combine PDFs for images belonging to the same system, we follow the hierarchical Bayesian procedure introduced by \cite{wan15,dah13}, which determines a combined \mbox{$P(z)$} from individual \mbox{$P_i(z)$} by accounting for the probability that each measured \mbox{$P_i(z)$} may be incorrect (\mbox{$p_{bad}$}). In short, the method inputs the individual \mbox{$P_i(z)$} if it is reliable, and uses a uniform \mbox{$P_i(z)$} otherwise. Then, assuming a flat prior in \mbox{$p_{bad}$} for \mbox{$p_{bad}\leq 0.5$}, we marginalize over all values of \mbox{$p_{bad}$} to calculate the combined \mbox{$P(z)$}. This method can introduce a small non-zero floor on the PDF, but this does not affect the peak in the distribution. 

\subsection{Weak Lensing Catalog}\label{wl}
Using ACS F814W observations of A370 from the HFF program, ellipticity measurements of 909 galaxies are identified as weak lensing constraints. To produce and reduce this catalog, we use the pipeline described by \cite{schr18a} which utilizes the \cite{erb01} implementation of the KSB+ algorithm \citep{kai95,lup97,hoe98} for galaxy shape measurements as detailed by \cite{schr07}. In addition, the pipeline employs pixel-level correction for charge-transfer inefficiency from \cite{mass14} as well as a correction for noise-related biases, and does temporally and spatially variable ACS point-spread function (PSF) modeling using the principal component analysis described by \cite{schr10}. \cite{schr18b} and Hern{\'a}ndez-Mart{\'{\i}}n et al. (in prep.) have extended earlier simulation-based tests of the employed shape measurement pipeline to the non-weak shear regime of clusters (for \mbox{$|g|\le 0.4$} where g is shear), confirming that residual multiplicative shear estimation biases are small (\mbox{$|m|\lesssim 5\%$}). Weak lensing galaxies extend to the edge of the ACS field of view and are individually assigned a photometric redshift from the ASTRODEEP photometry catalogs discussed in Section \ref{wl}. Individual redshifts are used for all galaxies in the catalog as constraints on the lens model. The weak lensing catalog is publicly available along with the lens model products on the HFF archive.\footnote{https://archive.stsci.edu/pub/hlsp/frontier/abell370/models/}

\begin{figure*}
\centering
    \begin{subfigure}
        \centering
        \includegraphics[width=8cm]{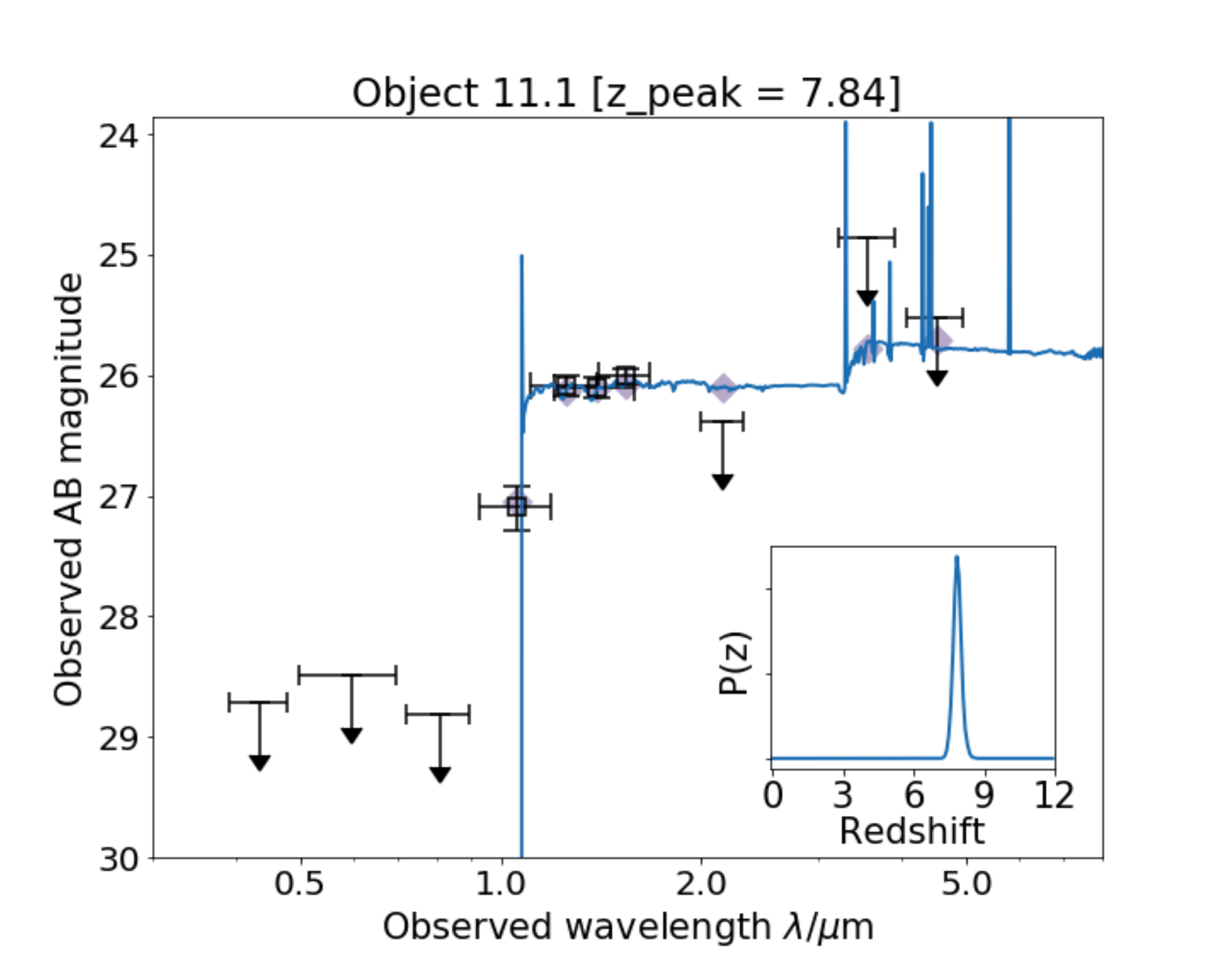}
    \end{subfigure}\quad
    \begin{subfigure}
        \centering
        \includegraphics[width=8cm]{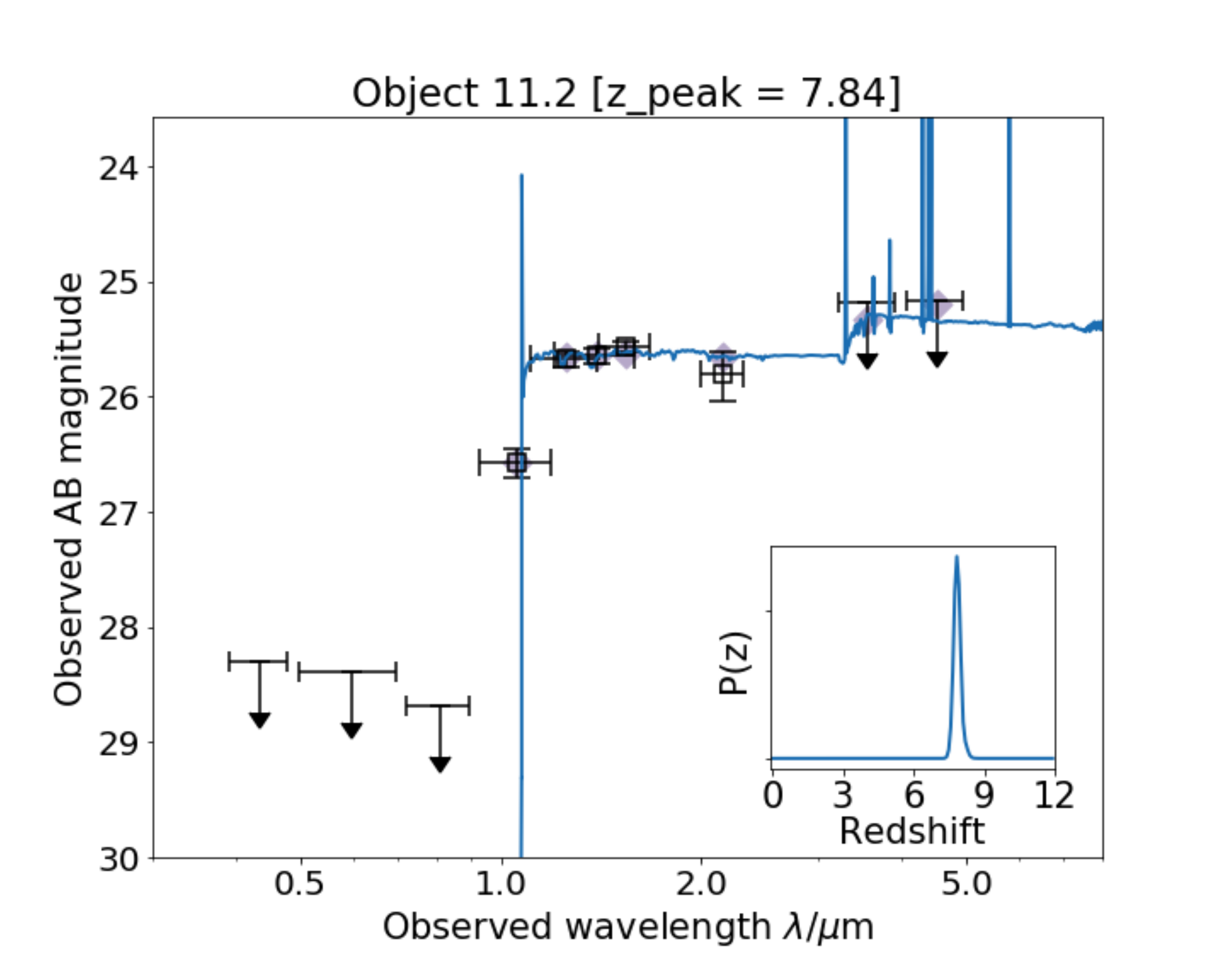}
    \end{subfigure}
    \caption{The SEDs and best-fit template for the multiple images in system 11. The IRAC fluxes were extracted from T-PHOT, and the SED fitting was done with the redshift-fitting code EAZY as described in Section \ref{phot_section}. Error bars and upper limits shown are 1-$\sigma$. The combined PDFs of the multiple images are shown in Figure \ref{811}.}
    \label{fig:mi1}
\end{figure*}

\subsection{Multiple Images}\label{mi_section}

Sets of multiple image candidates were visually inspected using HST color images by six independent teams in the HFF community, including ours, and are ranked based on the availability of a spectroscopic redshift and similarity of the images in color, surface brightness, and morphology. Six independent teams inspect and vote on each image on a scale of 1-4, 1 meaning the image has a secure redshift and 4 meaning the redshift measurement is poor and the image is difficult to visually associate with a system. Votes are averaged to represent the quality of the image. In this paper we use only systems containing a majority of images with an average score of 1.5 or less. This translates to multiply imaged systems that either have a spectroscopic redshift for each image in the system, or images that have PDFs in agreement to 1-\mbox{$\sigma$}. Alternatively, the system has at least one spectroscopically confirmed image and other images have convincingly similar colors, morphologies, and surface brightnesses. Our numbering scheme is adopted from \cite{lag17}; of our 39 multiply-imaged systems, 37 are spectroscopically confirmed (systems labeled ``z-spec" in Table \ref{tbl-2}, blue points in Figure \ref{fig:cc_comp}). 

These systems' spectroscopic redshifts have been collected over time, starting with systems 1 \citep{kne93}, 2 \citep{sou87}, and 3 \citep{rich14}. These, in addition to 10 unconfirmed systems, were used by \cite{rich14}. With GLASS spectroscopy, \cite{die18} confirmed these as well as systems 4, 6, 9, and 15. Finally, \cite{lag17} confirmed 10 additional systems (5, 7, 14, 16, 17, 18, 19, 20, 21, and 22). Following the lead of \cite{die18} and \cite{lag17}, we treat system 7 (named systems 7 and 10 in \cite{lag17} and systems 7 and 19 in \cite{die18}) as a single system due to the fact that all images appear to be from the same source galaxy at the same spectroscopic redshift (measured by \citealp{lag17}). We use all 39 systems as constraints in our model, including one that is lensed by a smaller cluster member on the outskirts of the field (system 37) and two others that are not spectroscopically confirmed (systems 8 and 11; see Figure \ref{811}). This is summarized in Table \ref{tbl-2}.

\subsection{System 11}
We note in particular system 11, a set of sources we believe to be multiply imaged, with photometric redshifts both peaking at \mbox{$z =7.84\pm0.02$} (Figure \ref{fig:mi1}). This system was found to be at \mbox{$z=5.9$} in \cite{rich14,die18}, and \mbox{$z=4.66$} in \cite{lag17}. In previous versions of our model, system 11 was found to be at \mbox{$z \sim 4$}. This redshift was obtained from HST only photometry, which has since been improved to include better ICL subtraction and \emph{Spitzer}/IRAC fluxes, as described in Section \ref{phot_section}. The photometric redshift of both images are now preferred at \mbox{$z =7.84\pm0.02$}. For a multiply imaged system such as system 11, which contains two images of opposite parity and similar surface brightnesses, the critical curve should appear between the images, approximately equidistant from each. Based on the critical curve placement near system 11, the new redshift is in broad agreement with all models of A370, and these results are consistent with photometric results presented by \cite{ship18}. 

We show the SED and best-fit template from EAZY in Figure \ref{fig:mi1}, where all error bars and upper limits shown are 1-$\sigma$. In object 11.1, the covariance index is found to be \mbox{$\sim$ 1.24} for both IRAC channels. The covariance index is defined as the ratio between the maximum covariance of the source with its neighbors over its flux variance, which serves as an indicator of how strongly correlated the source's flux is with its closest or brightest neighbor. Generally, a high covariance index (\mbox{$>$ 1}) is associated with more severe blending and large flux errors \citep{laid07,merl15}, so we treat these fluxes with caution. Because of the high confidence in visual detection of the object and its multiple image, we include the flux upper limits in our SED fit. However, when EAZY is run without these flux values included, the best-fit SED template and \mbox{$z\sim8$} solution remains, with a slightly broader PDF. The combined photometric redshift probability distributions are shown in Figure \ref{811}. 

The unlensed absolute magnitudes of the images are \mbox{$-18.68^{+0.10}_{-0.08}$} and \mbox{$-18.16^{+0.07}_{-0.08}$} for 11.1 and 11.2, respectively, where photometric errors in AB flux measurement and statistical errors in magnification are included. While these values are not in statistical agreement, the uncertainty in magnification close to the critical curve is larger than the statistical uncertainty in our model. While our model predicts positions of the sources well, we do not use brightness of sources as constraints. Ultimately, spectra will be needed to confirm or deny the redshift of the sources. Both images in system 11 fall outside of the coverage of the MUSE GTO program \citep{lag17}, but were observed by GLASS and with the Multi-Object Spectrometer for Infra-Red Exploration (MOSFIRE) instrument on Keck. However, these data do not constrain any noticeable spectroscopic features and therefore do not constrain the spectroscopic redshift (Hoag et al., in prep.).

While the images in system 11 are observed as relatively bright objects, they are intrinsically faint, which offers a unique chance to study a more representative example of a \mbox{$z\sim 8$} galaxy. The source being multiply imaged will allow for better statistics on the properties inferred about it. This makes the source an ideal target for \emph{James Webb Space Telescope}, as emission lines at this observed brightness will likely be detectable.

\subsection{Lens Modeling Procedure}
The lens modeling code used in this work, Strong and Weak Lensing United (SWUnited, \citealp{brad05,brad09}), uses an iterative \mbox{$\chi^2$} minimization method to solve for the gravitational potential on a grid. The method constructs an initial model assuming a range of profiles (we use the non-singular isothermal ellipsoid as our initial model here) and uses multiple images reconstructed in the source plane as constraints.  A \mbox{$\chi^2$} is calculated upon each iteration using gravitational potential values on a set of non uniform grid points on an adaptive grid. The grid uses higher resolutions near areas where there are many constraints and is determined by a set of user-created refinement regions, which consist of circles of given radii that appoint levels of resolution. We optimize the model using a \mbox{$\chi^2$} defined as:

\begin{equation}
    \chi^2 = \chi^2_\mathrm{SL} + \chi^2_\mathrm{WL} + \eta R,
\end{equation}
where \mbox{$\chi^2_{SL}$} is a strong lensing term in the source plane, \mbox{$\chi^2_{WL}$} is a weak lensing term that uses ellipticies of weakly lensed galaxies as constraints, and $\eta$ is a regularization parameter of the regularization function R that penalizes small-scale fluctuations in the gravitational potential. After finding a minimum $\chi^2$, the code produces convergence ($\kappa$), shear ($\gamma$), and magnification ($\mu$) from the best-fit solution. 

Our method differs from other parameterized codes in that we do not make any assumptions regarding light tracing mass. It is parameterized in that there are parameters which are obtained via minimization, i.e. the gravitational potential in each cell, but they are kept as general as possible and the minimization is done on a non-uniform grid, while other codes compare strong and weak lensing constraints in parameter space using a Bayesian approach and assuming simple parameterized models. In addition, we include weak lensing constraints that extend to the center of the cluster. While the method employed by \cite{die18} has the ability to use weak lensing constraints, they do not do so for A370, and no other groups from the HFF campaign use weak lensing constraints on this cluster.

\begin{figure*}[!!!!!!!ht]
\centering
    \begin{subfigure}
        \centering
        \includegraphics[width=8cm]{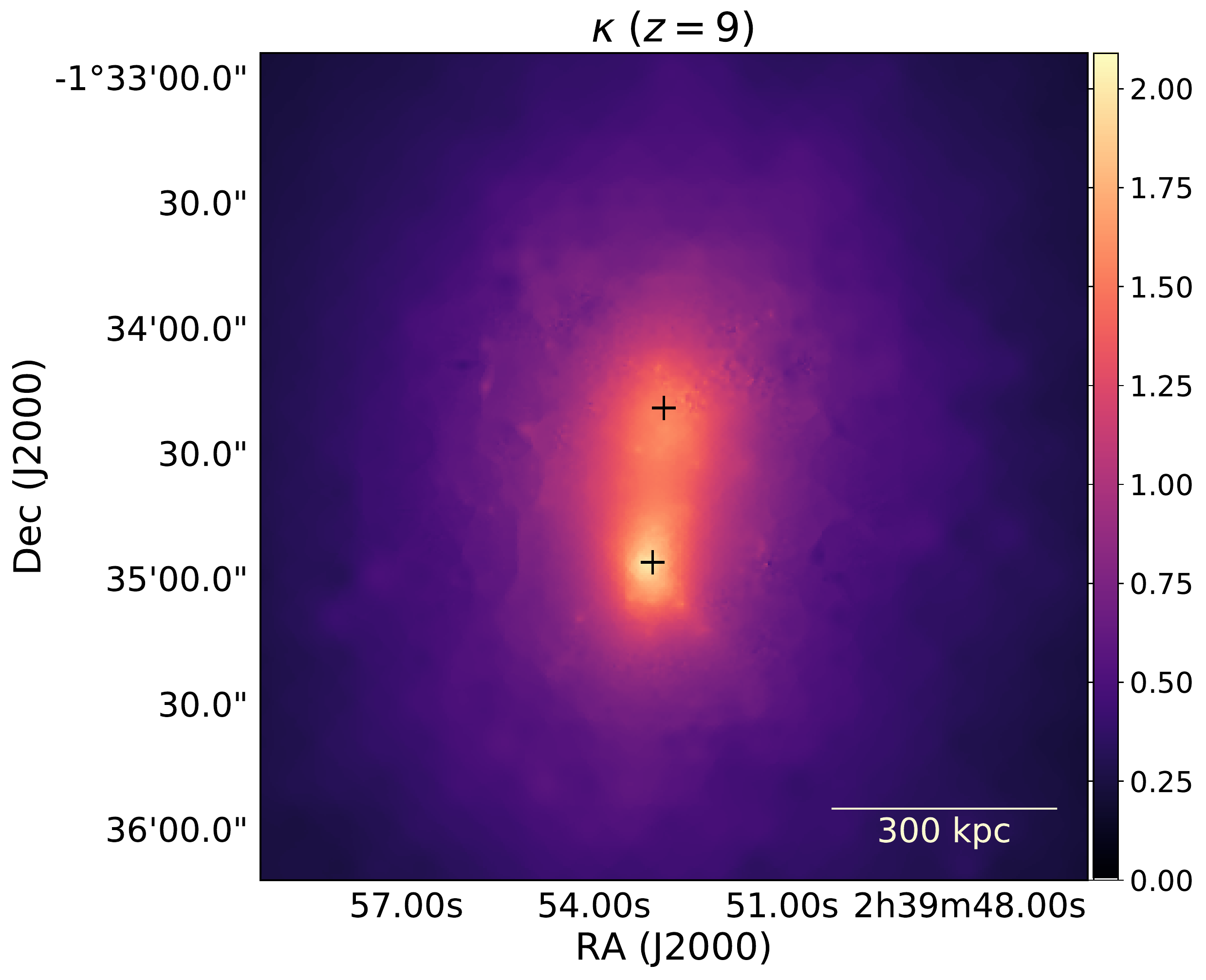}
    \end{subfigure}\quad
    \begin{subfigure}
        \centering
        \includegraphics[width=8cm]{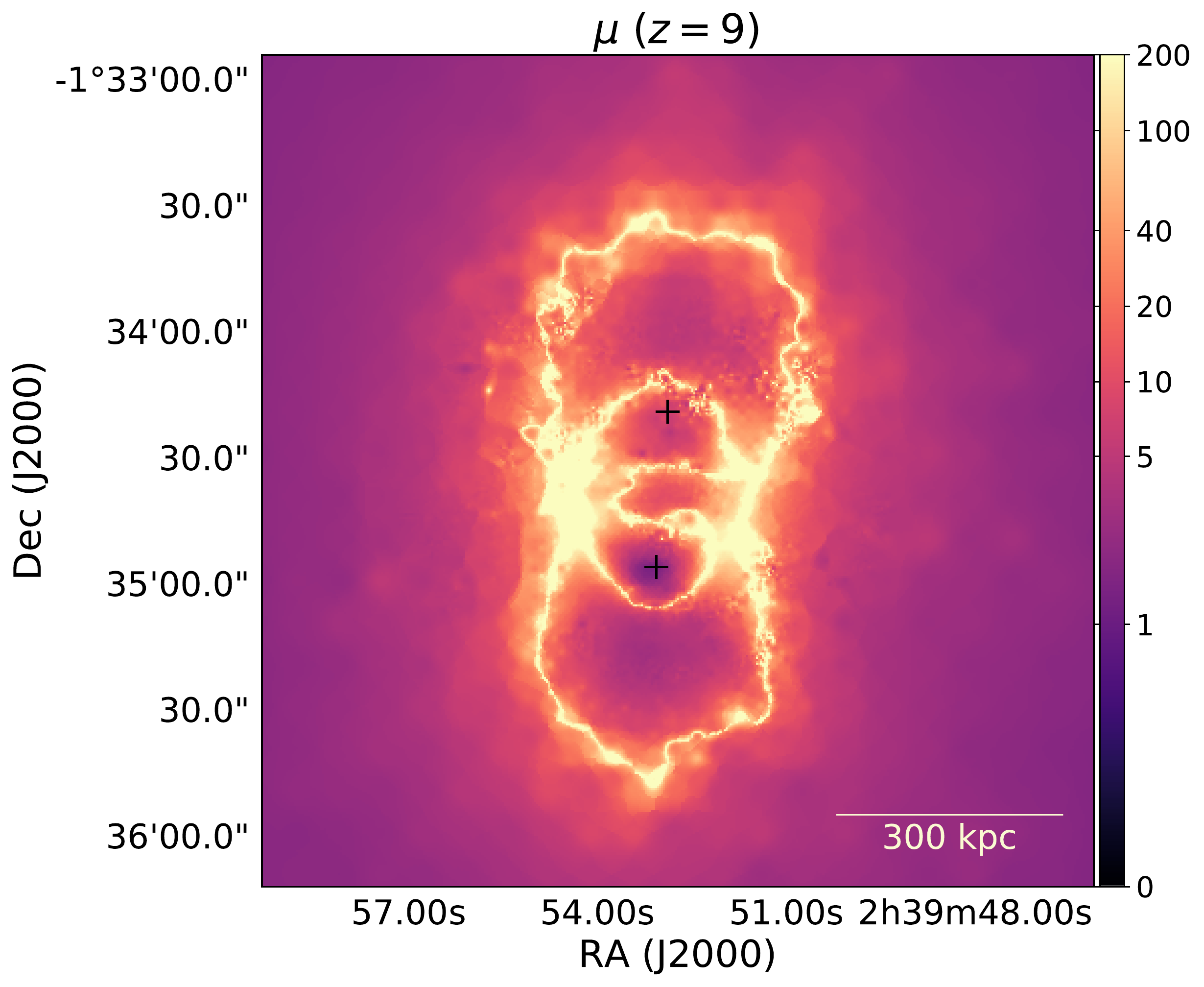}
    \end{subfigure}
\caption{\textbf{Left:} Convergence ($\kappa$) map of Abell 370 produced by our lens model for a source at \mbox{$z=9$}. BCG centers are shown as crosses. Two dominant peaks in mass density near the location of the BCGs are shown, with a small offset between the southernmost BCG and mass density peak. \textbf{Right:} Magnification ($\mu$) map of Abell 370 for a source at \mbox{$z=9$}. The yellow contour corresponds to maximum magnification values; a highly elliptical and extended critical curve is revealed, similar in shape to those found by other groups (e.g., \cite{die18,lag17} and other groups on the HFF archive \footnote{https://archive.stsci.edu/pub/hlsp/frontier/abell370/models/}). Orientation is the same as in Figure \ref{fig:cc_comp}.}
\label{fig:mag}
\end{figure*}
\subsection{Stellar Mass Map}
Rest-frame K-band flux has been shown to estimate stellar mass well due to its insensitivity to dust within the observed cluster \citep{bell03} and lack of dependence on star formation history \citep{kau98}. Since IRAC channel 1 (3.6$\mu$m, [3.6] hereafter) is the closest band corresponding to rest-frame K-band of the cluster, we use it to estimate stellar mass of A370 using flux in cluster members in this channel.
Cluster members are selected using the red sequence (F435W and F814W magnitudes), visually inspected to remove the obvious outliers, and redshifts are verified to be within \mbox{$\pm0.1$} of the mean cluster redshift (\mbox{$z=0.375$}) with GLASS spectroscopy. 

Following the procedure described by \cite{hoa16}, we create a mask of cluster members from an F160W segmentation map of the field, convolve the map with the IRAC channel 1 PSF, and resample onto the IRAC pixel grid. We then apply this mask to the IRAC channel 1 image in order to get a [3.6] map containing only light (to a good approximation) from cluster members. After smoothing the IRAC surface brightness map with a Gaussian kernel of \mbox{$\sigma=3$} pixels, we calculate luminosities of the cluster members using [3.6] flux, and apply a K-correction of -0.33 to bring them to K-band for the mean cluster redshift. We then multiply the map by a mass to light ratio, \mbox{$M_{\text{*}}/L = 0.95 \pm 0.26 M_{\odot}/L_{\odot}$}, obtained in \cite{bell03} assuming the diet Salpeter IMF. This choice contains 70$\%$ the mass of the Salpeter IMF for the same photometry, and is used here for comparison of our results to previous results \citep{wan15,hoa16,finney18}. 

Since IMF can change $\langle f\text{*} \rangle$ by as much as 50\%, this choice introduces our largest error in estimating stellar mass. Additional sources of error include our calculation of stellar mass using a single mass to light ratio and choice of template used to calculate the K-correction instead of deriving stellar mass from SED fitting. When comparing stellar mass calculations of both methods in clusters similar to A370, we find that this choice produces a 0.05 dex bias, which translates to a \mbox{$10\%$} underestimate in stellar mass using our method. Other errors include statistical errors and an underestimation of stellar mass due to not accounting for stars in the ICL. \cite{mon18} found A370 to have \mbox{$4.9 \pm 1.7\%$} of total light within a radius of \mbox{$R_{500}$} residing in the ICL. However, these errors are all sub-dominant and negligible compared to the uncertainty related to the choice of IMF \citep{bur15}.

\section{Results}

\subsection{Mass and Magnification}\label{massmag}
Convergence $\kappa$ and magnification $\mu$ maps for a source at \mbox{$z = 9$} are shown in Figure \ref{fig:mag}, displaying two dominant peaks. The southernmost brightest cluster galaxy (BCG) is roughly aligned with the convergence peak, however the northernmost $\kappa$ peak is significantly less concentrated and shows a small offset from the stellar mass. There are less significant peaks in the $\kappa$ map around the cluster members in the northeast and a bright cluster member in the southwest. The yellow contour in the magnification map is the critical curve, where the magnification is at a maximum. Magnification reaches up to $\mu \sim$ 10-20 within 1-2 arcseconds from the critical curve, while typical values of magnification range from $\mu \sim$ 2-5 near the edges of the HST field. 

\begin{figure*}
    \centering
    \includegraphics[width=17cm]{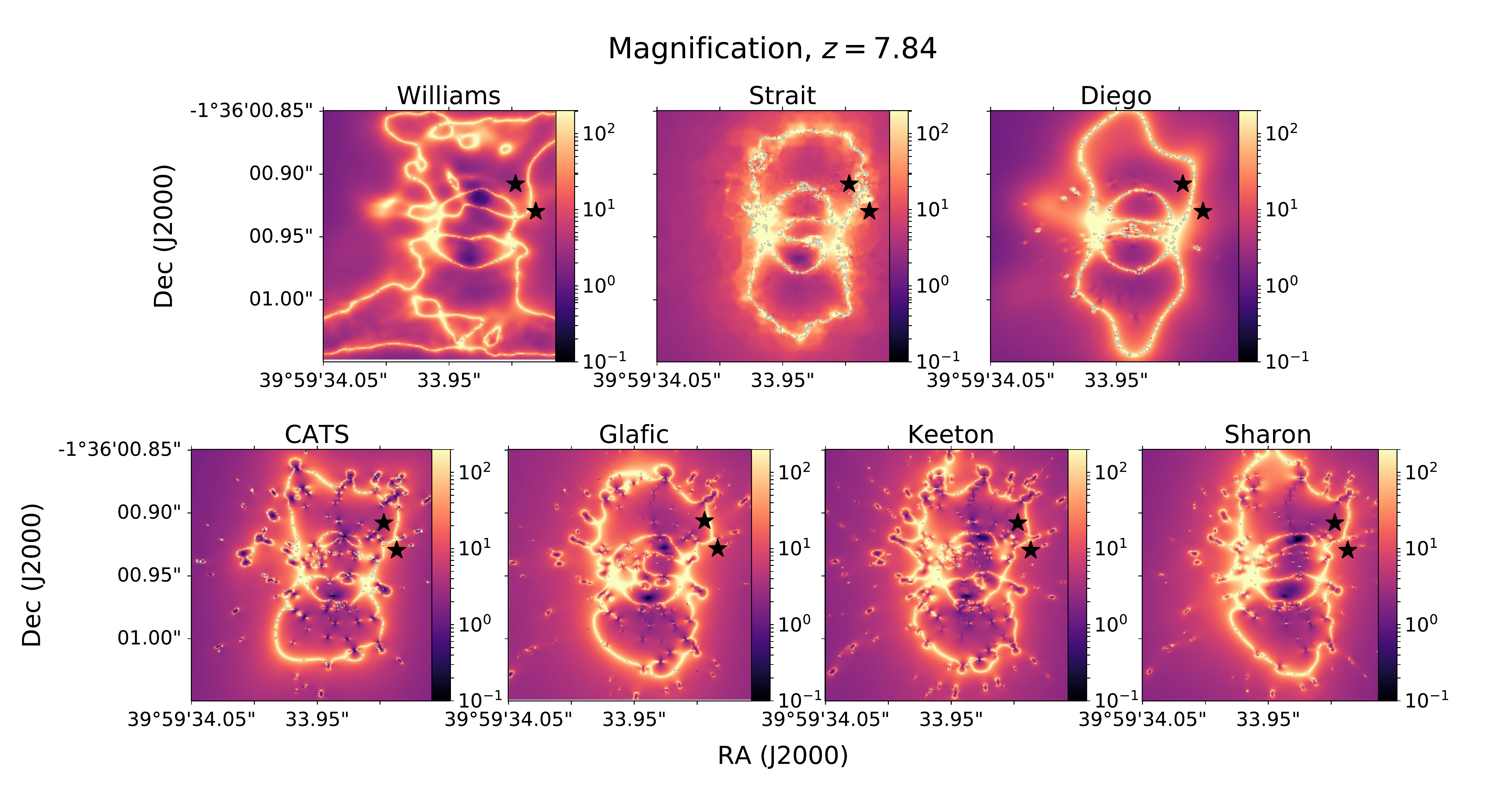}
    \caption{Critical curves at \mbox{$z=7.84$}, the photometric redshift of the images in System 11. Black stars mark the position of the multiple images. The CATS and Sharon groups use lenstool, which uses individual galaxies and other large parameterized mass components. Keeton and Glafic teams are also parametric models, and Diego uses a non-parametric lensing code but a light traces mass assumption. The Williams team along with ours make no such assumptions. All teams but ours used strong lensing constraints only. While shapes of critical curves vary, all groups, including ours (see Figure \ref{fig:cc_comp}), show good agreement for the images at \mbox{$z=7.84$}.}
    \label{fig:mags}
\end{figure*}

In the absence of an ability to compare our model to truth, a comparison of parametric, free-form, and grid-based modeling codes is helpful to properly account for the systematic uncertainties of each method that can produce this spread. When comparing our magnification map to previous models of A370, we only compare to models updated since the last data release. Group names are: Glafic \citep{ogu10,kaw17}, CATS \citep{lag17}, Diego \citep{die18}, Keeton, Merten, Sharon, and Williams. More information about each method can be found on the HFF archive\footnote{ https://archive.stsci.edu/pub/hlsp/frontier/abell370/models/}. As shown in Figure \ref{fig:mags}, our critical curves are approximately of the same ellipticity and extent, with a larger radial region compared to many of the groups. With the exception of the Williams map which has a boxy shape, the overall shapes are comparable. The critical curve at \mbox{$z=7.84\pm0.02$} (the redshift of System 11) for our model is shown in Figure \ref{fig:cc_comp} and at \mbox{$z=9$} in Figure \ref{fig:mag}. On smaller scales, the magnification levels differ greatly from group to group, particularly very close to the critical curves. The black stars in Figure \ref{fig:mags} correspond to the multiple images in System 11, and we find that the critical curves of all models fall in a reasonable place to be consistent with the new redshift. Explicitly, \cite{lag17} finds that model constraints allow a range of \mbox{$2.5 < z < 10$} when the redshift of this system is varied as a free parameter, consistent with a \mbox{$z=7.84\pm0.02$} solution. 

\begin{figure*}
    \centering
    \includegraphics[width=15cm]{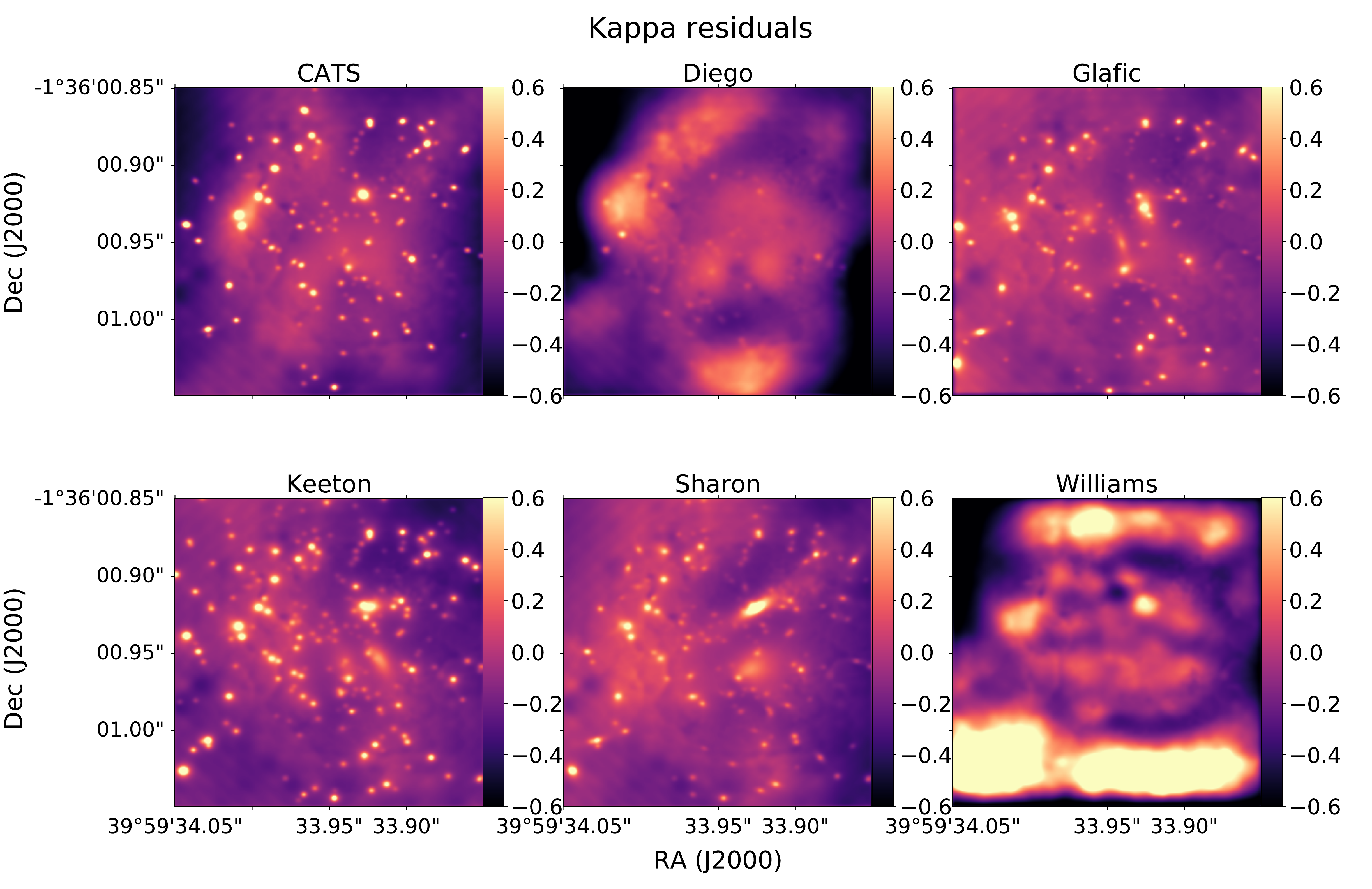}
    \caption{Kappa residuals (\mbox{$(\kappa_i - \kappa_{VS})/\kappa_{VS}$}), where \mbox{$\kappa_i$} are convergence maps from six models and \mbox{$\kappa_{VS}$} is the convergence map presented here, smoothed with a Gaussian filter of \mbox{$\sigma=10$}. See caption of Figure \ref{fig:mags} and Section \ref{massmag} for a description of the groups and lens modeling methods.}
    \label{fig:kappa_resids}
\end{figure*}
In Figure \ref{fig:kappa_resids} we compare surface mass density (\mbox{$\kappa$}) distributions. There are obvious differences, such as clear high residuals over cluster members in the groups who use lensing codes that assume light traces mass (Sharon, CATS, Glafic, Keeton, Diego). There is also a large residual in the south of the Williams map, where the Williams model differs from most other models. Compared to \cite{die18} and \cite{lag17}, we have smaller \mbox{$\kappa$} values in the northeast.

\subsection{Stellar to Total Mass Ratio}
\begin{figure*}
\centering
\begin{subfigure}
    \centering
    \includegraphics[width=8cm]{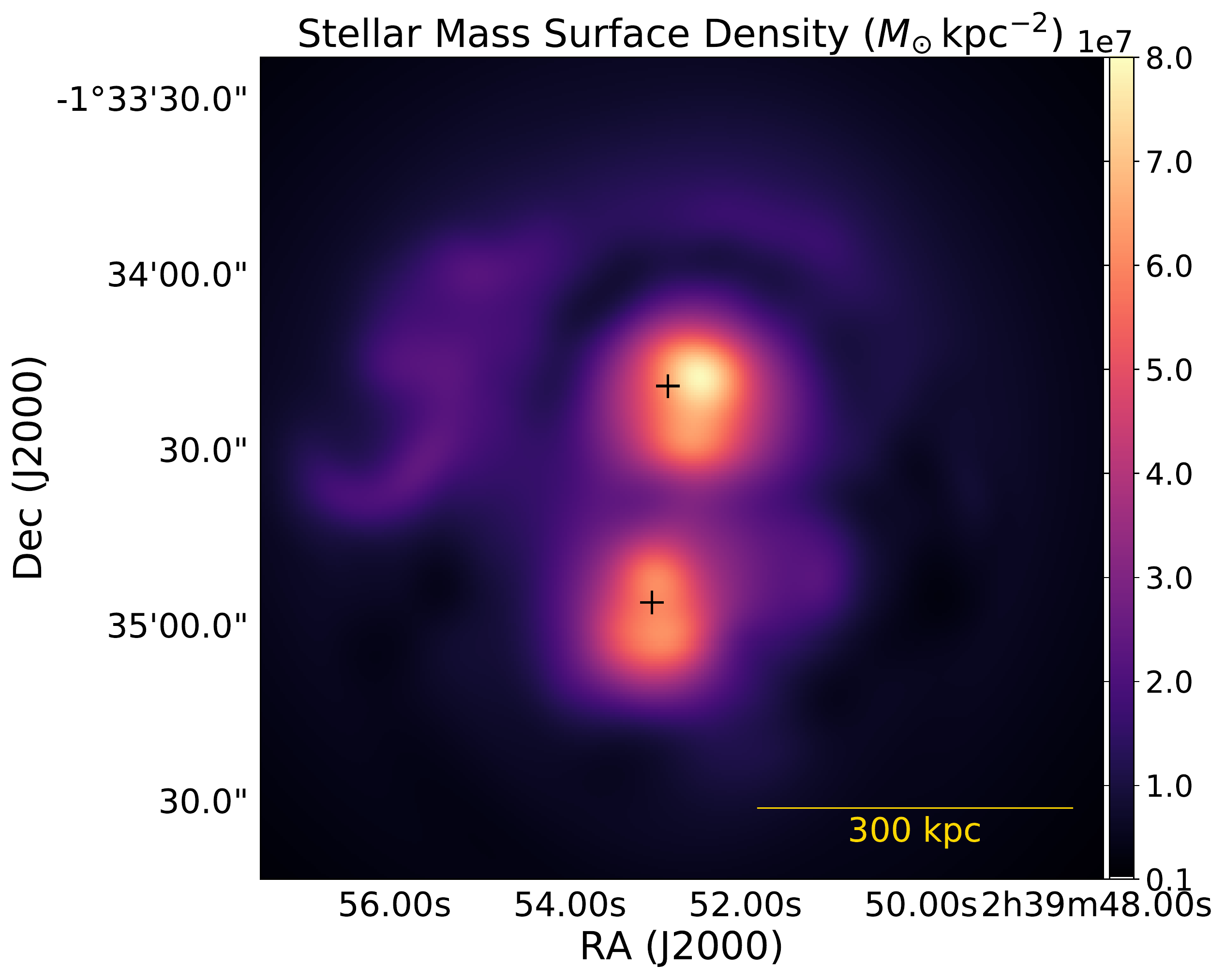}
\end{subfigure}
\begin{subfigure}
    \centering
    \includegraphics[width=8cm]{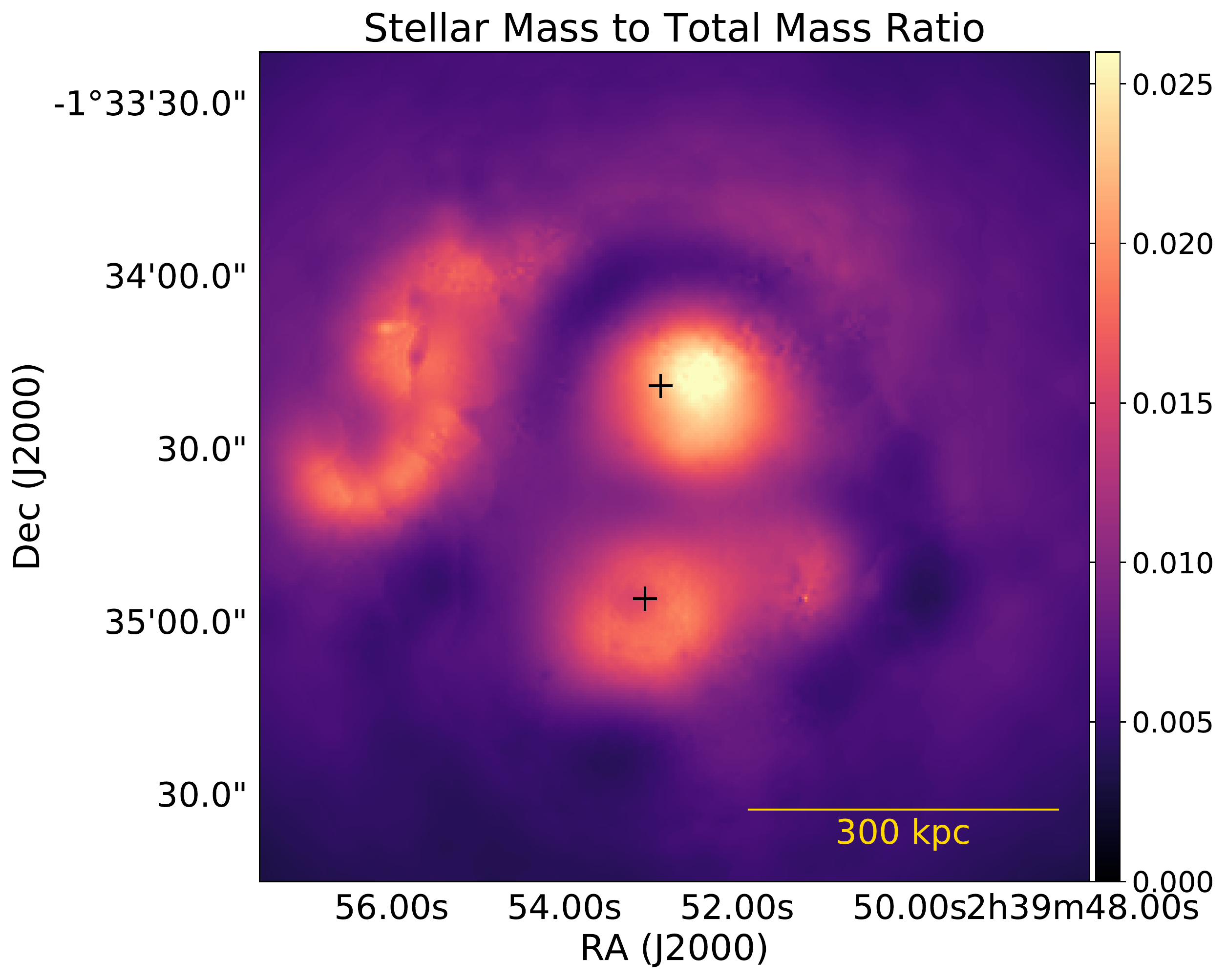}
\end{subfigure}
\caption{\textbf{Left:} Stellar surface mass density in units of $M_{\odot}$ kpc$^{-2}$, produced from an IRAC [3.6] image as described in Section 3.4. \textbf{Right:} Stellar to total mass ratio (\mbox{$f\text{*}$}), produced by matching the resolution of the stellar mass map to the adaptive grid from the total mass map and dividing. The center of BCGs are shown as crosses. The largest peak in this figure is over the northernmost BCG where there are high values of stellar mass due to a bright BCG and low values of total mass, from the lens model. The black crosses are centered over the peak in light from each BCG as determined from the HST image.}
\label{f*}
\end{figure*}

To study the difference in stellar mass from cluster members and total cluster mass, we look at the stellar mass to total mass fraction, \mbox{$f\text{*}$}. We obtain an \mbox{$f\text{*}$} map by dividing the total stellar mass density in a 0.3 Mpc radius by the total projected mass density in the same radius, after adjusting the resolution (i.e. by smoothing) and pixel scale of the stellar mass density map to match that of the total mass density map, which was determined by the refinement region discussed in Section 3 (see \cite{hoa16} for details on this procedure). The resulting \mbox{$f\text{*}$} map is shown in Figure \ref{f*}. There is considerable variation throughout the map, reaching values near 0.03-0.04 on top of the northern BCG. The stellar mass and \mbox{$f\text{*}$} map reflects what is expected, with higher values around the cluster members in the northeast and to the west over a particularly bright galaxy. There is a peak in stellar mass on top of both BCGs, as expected, but the northernmost peak is higher and offset by a modest amount from the stellar mass peak caused by the BCG. The high offset peak in combination with a less significant peak in the total mass on the northern BCG creates the highest peak in the \mbox{$f\text{*}$} map. 

We find that average \mbox{$f\text{*}$} in a circular aperture of radius 0.3 Mpc is \mbox{$\langle f\text{*} \rangle = 0.011 \pm 0.003$}, when centered over a midpoint in between the BCGs. We select BCG centers using flux peaks in F160W images, however we cannot include details of how centers were chosen in other analyses presented here, as that information was not publicly available. If re-calculated using a radius of the same size centered on the southern and northern BCG, we find a value of \mbox{$\langle f\text{*} \rangle = 0.011 \pm 0.003$} and \mbox{$\langle f\text{*} \rangle = 0.012 \pm 0.003$}, respectively. As was the case with the stellar mass map, the choice of IMF is the largest source of error by an order of magnitude, with the ability to change our value of \mbox{$\langle f\text{*}\rangle$} by as much as 50\%. 

In comparing our average value of stellar to total mass to clusters of similar mass and redshift, we find good agreement.  Average $f\text{*}$ obtained for a radius of 0.3 Mpc around MACSJ0416 (\mbox{$z=0.396$}) in \cite{hoa16} is \mbox{$\langle f\text{*} \rangle = 0.009 \pm 0.003$}, and \cite{finney18} obtain a value of \mbox{$\langle f\text{*}\rangle = 0.012^{+0.003}_{-0.005}$} for MACS1149 (\mbox{$z=0.544$}). Both calculations use SWUnited maps and a diet Salpeter IMF. Similarly, using the SWUnited maps produced by \cite{wan15} for Abell 2744 (\mbox{$z=0.308$}), we find a value of \mbox{$\langle f\text{*}\rangle = 0.003 \pm 0.001$}. In another analysis of MACS0416, \cite{jau16} find a value of \mbox{$\langle f\text{*}\rangle = 0.0315 \pm 0.0057$} using a Salpeter IMF and a radius of 200 kpc.  When re-calculated using the diet Salpeter IMF, we get a value of $0.0221 \pm 0.0057$ for MACSJ0416. In a study of 12 clusters near \mbox{$z \sim 0.1$} with masses greater than \mbox{$2\times 10^{14} M_{\odot}$}, \cite{gon13} found a value of \mbox{$\langle f\text{*}\rangle$} to be 0.0015-0.005 in a radius of \mbox{$1.53\pm0.08$} Mpc. \cite{bah14} find a similar value of 0.010 $\pm$ 0.004 on all scales larger than 200 kpc, when examining \mbox{$f\text{*}$} for more than 13,823 clusters in the redshift range \mbox{$0.1 < z < 0.3$}, selected from the MaxBCG catalog \citep{koe07}; however, they use SDSS i-band to calculate stellar mass and assume a Chabrier IMF. When re-calculated using a diet-Salpeter IMF, we obtain \mbox{$f\text{*}=0.012 \pm 0.005$} for this sample. These results are summarized in Table \ref{tbl-1}.
\begin{deluxetable*}{llllll}
\tablecaption{\label{tbl-1} Comparison of \mbox{$\langle f\text{*}\rangle$} to values in literature}
\tablewidth{0pt}
\tablehead{
\colhead{$\langle f\text{*} \rangle$} & \colhead{Object} & \colhead{Redshift} & \colhead{Radius}  & \colhead{IMF} & \colhead{Reference}
}
\startdata
0.011 $\pm$0.003 & A370 & 0.375 & 0.3 Mpc & diet Salpeter & This paper\\
0.009 $\pm$ 0.003 & MACS0416 & 0.396 & 0.3 Mpc & diet Salpeter & \cite{hoa16}\\
0.012$^{+0.003}_{-0.005}$ & MACS1149 & 0.544 & 0.3 Mpc & diet Salpeter & \citealp{finney18}\\
0.0015-0.005 & 12 clusters, \mbox{$z\sim0.1$}  & \mbox{$\sim 0.1$} & \mbox{$1.53\pm0.08Mpc$} & Salpeter & \cite{gon13}  \\
0.003 $\pm$ 0.001 & A2744 & 0.308 & 0.3 Mpc & diet Salpeter & \cite{wan15} \\
0.0221 $\pm$ 0.0057 & MACS0416 & 0.396 & 200 kpc & Salpeter & \cite{jau16}\\
0.010 $\pm$ 0.004 & clusters from MaxBCG & \mbox{$0.1<z<0.3$}& \mbox{$>200$} kpc & Chabrier & \cite{bah14}\\

\end{deluxetable*}

In general, we find that stellar mass traces total mass in the center of the cluster reasonably well, with the exception of a small offset near the northern BCG. This could be due to the cluster's bimodal distribution which indicates a possible merger. In comparing to the smoothed light maps presented in \cite{lag17}, we see a similar distribution over each BCG and the ``crown" of galaxies in the north. Our total mass over those cluster members, however, is lower than theirs, as seen in the peak in the northeast of the \mbox{$f\text{*}$} map (Figure \ref{f*}). This is also reflected in the positive residual seen in the CATS panel of Figure \ref{fig:kappa_resids}. 

The differences in total mass in the central part of the cluster mentioned in Section \ref{massmag} also appear in the critical curve placement, which can be seen in Figure \ref{fig:cc_comp}. The critical curve shown in Figure \ref{fig:cc_comp} is for \mbox{$z\sim 7.84$} but crosses the radial system 7 (\mbox{$z=2.75$}), meaning there is likely a higher \mbox{$\kappa$} and a larger critical curve in that region. This results in a lack of ability to recreate the system 7 images, but can be improved upon by adding a peaky mass clump to the center of the lens model. \citealp{die18} showed that adding a mass clump respresenting stellar mass improves their model in the central region of the cluster, where system 7 resides. Similarly, this addition of mass in the center of the two BCGs improves our model's prediction of system 7 images to sub-arcsecond precision. While \cite{die18} find that this addition of mass causes high $f*$ values ($30-100\%$) in the center of the BCGs, our model predicts a more moderate value of \mbox{$f* \sim 3 \%$}.


\section{Conclusions}
A370 is a cluster located at \mbox{$z = 0.375$} that acts as a powerful gravitational lens, behind which deep HST images have revealed 39 multiply-imaged source galaxies consisting of 114 images. Using spectroscopic redshifts from MUSE and GLASS, we produce a total projected mass density and magnification map with the grid-based lens modeling code SWUnited. Using IRAC [3.6] images, we calculated stellar mass and \mbox{$f\text{*}$}, the stellar to total mass density fraction in the field. Our main results are as follows:

\begin{enumerate}
    \item Using 37 multiply-imaged systems that have spectroscopically confirmed redshifts and 2 systems with photometric estimates, we constrain the total mass density distribution using lens modeling code SWUnited. The convergence ($\kappa$) and magnification ($\mu$) maps that we produce broadly agree with other models in ellipticity and size. There are discrepancies in the exact placement of the critical curves, producing variations in magnification levels in regions of high magnification, but there is agreement elsewhere. Our $\kappa$ map shows two dominant peaks, the northernmost peak being more diffuse than that of the southern BCG.
    \item There is a multiply imaged system consisting of two images, both with photometric redshifts peaking at \mbox{$z=7.84\pm0.02$}. This result comes from ASTRODEEP-like photometry and is in agreement with photometry by \cite{ship18}. The unlensed absolute magnitudes of the images are not consistent with that of an individual multiply imaged source at the 3-\mbox{$\sigma$} level, when considering only statistical errors. In order to be consistent with the same source, an additional systematic must be present, most likely in the magnification ratio.
    \item The \mbox{$f\text{*}$} map that we produce using an IRAC 3.6$\mu$m image shows considerable variation throughout the field of view. There are expected dominant peaks in the \mbox{$f\text{*}$} map near the BCGs, with higher values in the northern BCG. This is due to a modest offset in total mass and stellar mass in the northern part of the cluster, possibly due to a merger process.
    \item We obtain a value of stellar to total mass ratio within 0.3 Mpc of \mbox{$f*= 0.011 \pm 0.003$} (stat.), with the largest systematic error due to our choice of IMF. This value agrees broadly with clusters of similar size and redshift, and with values found for large scale ($>$ 1 kpc) cluster environments.
\end{enumerate} 
\section*{Acknowledgements}
Support for this work was provided by NASA through an award issued by JPL/Caltech (for SURFS UP project) and by HST /STScI HST-AR-13235, HST-AR-14280, and HST -GO-13177. Support for the Grism Lens-Amplified Survey from Space (GLASS) (HST-GO-13459) was provided by NASA through a grant from the Space Telescope Science In- stitute (STScI). This work utilizes gravitational lensing models produced by PIs Bradac, Richard, Natarajan \& Kneib (CATS), Sharon, Williams, Keeton, and Diego. This lens modeling was partially funded by the HST Frontier Fields program conducted by STScI. STScI is operated by the Association of Universities for Research in Astronomy, Inc. under NASA contract NAS 5-26555. Lens models were obtained from the Mikulski Archive for Space Telescopes (MAST). B.V. acknowledges the support from an  Australian Research Council Discovery Early Career Researcher Award (PD0028506). T.T. and X.W. acknowledge support by NASA through HST grants HST-GO-13459 and HST-GO-14280.

\clearpage
\LongTables
\begin{deluxetable*}{cccccc}
\tabletypesize{\footnotesize}
\tablecaption{\label{tbl-2} Multiply imaged arc systems behind A370 \\ \citealp{lag17} = L17 \citealp{die18} = D18 \\ Lagattuta, in prep. = L18} 
\tablewidth{0pt}
\tablehead{
\colhead{ID} & \colhead{RA} & \colhead{Dec}  & \colhead{\mbox{$z_{used}$}} & \colhead{Quality Flag}& \colhead{Reference}
}

\startdata
1.1 & 39.976290 & -1.576023 & 0.8041 & 3 & D18,L17 \\
1.2 & 39.967161 & -1.576876 & 0.8041 & 3 & D18,L17 \\
1.3 & 39.968546 & -1.576618 & 0.8041 & 3 & D18,L17 \\
\hline \\
2.1 & 39.973789 & -1.584241 & 0.7251 & 3 & D18,L17\\
2.2 & 39.970973 & -1.585035 & 0.7251 & 3 & D18,L17\\
2.3 & 39.968741 & -1.584507 & 0.7251 & 3 & D18,L17\\
2.4 & 39.969560 & -1.584804 & 0.7251 & 3 & D18,L17\\
2.5 & 39.969560 & -1.584804 & 0.7251 & 3 & D18,L17\\
\hline \\
3.1 & 39.978943 & -1.5674553 & 1.9553 & 3 & D18,L18\\
3.2 & 39.968493 & -1.565796 & 1.9553  & 3 & D18,L18\\
3.3 & 39.965668 & -1.566849 & 1.9553  & 2 & D18,L18\\
\hline \\
4.1 &39.979607  &-1.576288 & 1.2728  & 3 & D18,L17\\
4.2 & 39.970725 & -1.576203 & 1.2728 & 3 & D18,L17\\
4.3 & 39.961928 & -1.577890 & 1.2728 & 3 & D18,L17\\
\hline \\
5.1 & 39.973486 & -1.589050 & 1.2774 & 3 & L17\\
5.2 & 39.971018 & -1.589217 & 1.2774 & 3 & L17\\
5.3 & 39.969130 & -1.589053 & 1.2774 & 3 & L17\\
\hline \\
6.1 & 39.969445 & -1.577200 & 1.063 & 3 & D18,L17\\
6.2 & 39.964328 & -1.578246 & 1.063 & 3 & D18,L17\\
6.3 & 39.979593 & -1.577109 & 1.063 & 3 & D18,L17\\
\hline \\
7.1 & 39.969775 & -1.5804306 & 2.7512 & 3 & L17\\
7.2 & 39.969871 & -1.5807722 & 2.7512 & 3 & L17\\
7.3 & 39.968808 & -1.5856333 & 2.7512 & 2 & L17\\
7.4 & 39.986554 & -1.5775806 & 2.7512 & 2 & L18\\
7.5 & 39.961542 & -1.5800056 & 2.7512 & 1 & L18\\
7.6 & 39.968567 & -1.5717611 & 2.7512 & 3 & L18 (10.1) \\
7.7 & 39.968004 & -1.570875  & 2.7512 & 3 & L18 (10.2) \\
\hline \\
8.1 & 39.964471	& -1.569825 & 2.98 & photo-z \\
8.2 & 39.9619 & -1.5736389 & 2.98 & photo-z  \\
\hline \\
9.1 & 39.9624 & -1.5778861 & 1.5182 & 3 & D18,L18\\
9.2 & 39.969483 & -1.5762667 & 1.5182 & 2 & D18,L18\\
9.3 & 39.982017 & -1.5765333 & 1.5182 & 2 & D18,L18\\
\hline \\
11.1 & 39.963804 & -1.5693611 & 7.84 & photo-z & \citealp{rich14},\\
& & & & & D18 (\mbox{$z=5.93$}),\\
& & & & & L17 (\mbox{$z=4.66$})  \\
11.2 & 39.960771 & -1.5741472 & 7.84 & photo-z & \citealp{rich14},\\
& & & & & D18 (\mbox{$z=5.93$}),\\ 
& & & & & L17 (\mbox{$z=4.66$}) \\
\hline \\
12.1 & 39.984112 & -1.570880 & 3.4809 & 3 & L18\\
12.2 & 39.969622 & -1.566629 & 3.4809 & 3 & L18\\
12.3 & 39.959208 & -1.575238 & 3.4809 & 3 & L18\\
\hline \\
13.1 & 39.979521 & -1.571773 & 4.2467 & 3  & L18\\
13.2 & 39.975193 & -1.568811  & 4.2467 & 3 &  L18 \\
13.3 & 39.956753 & -1.5775058 & 4.2467 & 3 &  L18 \\
\hline \\
14.1 & 39.972283 & -1.5779833 & 3.1277 & 3 & L17 \\
14.2 & 39.972192 & -1.5801027 & 3.1277 & 3 & L17 \\
14.3 & 39.974183 & -1.5856083 & 3.1277 & 3 & L17 \\
14.4 & 39.981313 & -1.5781583 & 3.1277 & 3 & L18\\
14.5 & 39.957671 & -1.5804472 & 3.1277 & 3 & L18\\
\hline \\
15.1 & 39.971328 & -1.580604 & 3.7085  & 3 & L17\\
15.2 & 39.971935 & -1.5870512 & 3.7085 & 3 & L17\\
15.3 & 39.971027 & -1.5777907 & 3.7085 & 3 & L17\\
15.4 & 39.984017 & -1.5784514 & 3.7085 & 3 & L18\\
\hline \\
16.1 & 39.964016 & -1.5880782 & 3.7743 & 3 & L17\\
16.2 & 39.966037 & -1.5890355 & 3.7743 &  & L18\\
16.3 & 39.984414 & -1.5841111 & 3.7743 & 3 & L18\\
\hline \\
17.1 & 39.969758 & -1.5885333 & 4.2567 & 3 & L17 \\
17.2 & 39.985403 & -1.5808406 & 4.2567 & 3 & L18\\
17.3 & 39.960235 & -1.5836508 & 4.2567 & 3 & L18\\
\hline \\
18.1 & 39.97583  & -1.5870613 & 4.4296 & 3 & L17\\
18.2 & 39.981476 & -1.5820728 & 4.4296 & 3 & L18\\
18.3 & 39.957362 & -1.5820861 & 4.4296 & 3 & L18\\
\hline\\
19.1 & 39.971996 & -1.5878654 & 5.6493 & 3 & L17\\
19.2 & 39.985142 & -1.5790944 & 5.6493 & 3 & L18\\
19.3 & 39.958316 & -1.5813093 & 5.6493 & 3 & L18\\
\hline\\
20.1 & 39.965271 & -1.5878028 & 5.7505 & 3 & L17\\
20.2 & 39.963608 & -1.5868833 & 5.7505 & 3 & L17\\
\hline\\
21.1 & 39.966575 & -1.5846139 & 1.2567 & 3 & L17\\
21.2 & 39.967383 & -1.5850278 & 1.2567 & 3 & L17\\
21.3 & 39.981539 & -1.5814028 & 1.2567 & 3 & L18\\
\hline \\
22.1 & 39.974408 & -1.5861 & 3.1277 & 3 & L17\\
22.2 & 39.981675 & -1.5796861 & 3.1277 & 2 & L18\\
22.3 & 39.957906 & -1.5810108 & 3.1277 & 3 & L18\\
\hline \\
23.1 & 39.980254 & -1.5667639 & 5.9386 & 3 & L18\\
23.2 & 39.957314 & -1.572744  & 5.9386 & 3 & L18\\
23.3 & 39.977165 & -1.5662748 & 5.9386 & 3 & L18\\
\hline \\
24.1 & 39.963113 & -1.5705944 & 4.9153 & 3 & L18\\
24.2 & 39.962029 & -1.5723361 & 4.9153 & 3 & L18\\
\hline \\
25.1 & 39.987325 & -1.5788667 & 3.8084 & 3 & L18\\
25.2 & 39.96195  &-1.5831694  & 3.8084 & 3 & L18\\
25.3 & 39.966984 & -1.5867999 & 3.8084 & 2 & L18\\
\hline \\
26.1 & 39.979939 & -1.5713902 & 3.9359 & 3 & L18\\
26.2 & 39.974464 & -1.5680938 & 3.9359 & 3 & L18\\
26.3 & 39.957165 & -1.5769585 & 3.9359 & 3 & L18\\
\hline \\
27.1 & 39.972446 & -1.567157 & 3.0161 & 3 & L18\\
27.2 & 39.980694 & -1.571125 & 3.0161 & 3 & L18\\
27.3 & 39.95829  &-1.5759068 & 3.0161 & 3 & L18\\
\hline \\
28.1 & 39.963492 & -1.5822806 & 2.9112 & 3 & L18\\
28.2 & 39.967058 & -1.5845583 & 2.9112 & 3 & L18\\
28.3 & 39.987816 & -1.5774528 & 2.9112 & 3 & L18\\
\hline \\
29.1 & 39.9681 & -1.564825 & 4.4897 & 3 & L18\\
29.2 & 39.983551 & -1.5675575 & 4.4897 & 2 & L18\\
29.3 & 39.960908 & -1.5690194 & 4.4897 & 1 & L18\\
\hline \\
30.1 & 39.983459 & -1.5704492 & 5.6459 & 3 & L18\\
30.2 & 39.972404 & -1.5663533 & 5.6459 & 3 & L18\\
\hline \\
31.1 & 39.972404 & -1.5693301 & 5.4476 & 3 & L18\\
31.2 & 39.980667 & -1.5747346 & 5.4476 & 2 & L18\\
31.3 & 39.956158 & -1.5786786 & 5.4476 & 2 & L18\\
\hline \\
32.1 & 39.966286 & -1.5693446 & 4.4953 & 3 & L18\\
32.2 & 39.988098 & -1.5751871 & 4.4953 & 3 & L18\\
32.3 & 39.960682 & -1.5783795 & 4.4953 & 3 & L18\\
\hline \\
33.1 & 39.962723 & -1.5860035 & 4.882 & 3 & L18\\
33.2 & 39.966217 & -1.5879961 & 4.882 & 3 & L18\\
\hline \\
34.1 & 39.970108 & -1.5701499 & 5.2437 & 3 & L18\\
34.2 & 39.971806 & -1.5880395 & 5.2437 & 3 & L18\\
34.3 & 39.958565 & -1.5817008 & 5.2437 & 3 & L18\\
34.4 & 39.985046 & -1.579559  & 5.2437 & 3 & L18\\
\hline \\
35.1 & 39.981541 & -1.5658624 & 6.1735 & 3 & L18\\
35.2 & 39.975826 & -1.5644423 & 6.1735 & 3 & L18\\
\hline \\
36.1 & 39.962444 & -1.5807098 & 6.2855 & 3 & L18\\
36.2 & 39.965996 & -1.5843844 & 6.2855 & 3 & L18\\
\hline \\
37.1 & 39.97039 & -1.5687943 & 5.6489 & 3  & L18\\
37.2 & 39.970428 & -1.5694203 & 5.6489 & 3 & L18 \\
\hline \\
38.1 & 39.977154 & -1.5737917 & 3.1563 & 3 & L18\\
38.2 & 39.975071 & -1.5721161 & 3.1563 & 3 & L18\\
\hline \\
39.1 & 39.965442 & -1.5780222 & 1.2777 & 2 & L18\\
39.2 & 39.967933 & -1.5773472 & 1.2777 & 2 & L18\\
39.3 & 39.982296 & -1.576975  & 1.2777 & 1 & L18\\
\hline \\
40.1 & 39.963579 & -1.5656333 & 1.0323 & 3 & L18\\
40.2 & 39.962958 & -1.5661111 & 1.0323 & 3 & L18\\
40.3 & 39.963375 & -1.5659528 & 1.0323 & 3 & L18\\
\end{deluxetable*}
\bibliographystyle{apj}
\bibliography{A370_vstrait}

\end{document}